\documentclass{aa}
\usepackage{graphics}
\usepackage{times}
\begin{document}

\thesaurus{03(11.01.2; 11.17.3; 13.18.1)}

\title{Beaming models and the correlations between the core
dominance parameter and the core and extended powers of quasars}

\author{Chris Simpson\thanks{chris@naoj.org}}
\institute{Subaru Telescope, 650 N.~A`Oh\={o}k\={u} Place, Hilo, HI
96720, USA}

\date{Received 3 August 1998 / Accepted 20 August 1998}
\titlerunning{Beaming models and correlations in quasars}
\maketitle

\begin{abstract}
We investigate the recent claim by Qin et al.\ that the observed
correlations (or lack thereof) between the core dominance parameter
and the core and extended powers of samples of lobe- and
core-dominated quasars is in contradiction with beaming models.
Contrary to their conclusion, we find that their results are in
perfect agreement with such models, and support this assertion with
Monte Carlo simulations.
\end{abstract}

\section{Introduction}

It is widely believed that radio-loud quasars and powerful radio
galaxies (FR\,II-type according to the scheme of Fanaroff \& Riley
\cite{fanaroff}) differ only by the orientation of their radio axes
with respect to the observer's line of sight (e.g.\ Barthel
\cite{barthel}). A related scenario proposes that the cores of these
radio sources are dominated by beamed emission which can be
significantly enhanced by Doppler boosting, while the extended
emission is isotropic (e.g.\ Orr \& Browne \cite{orr}). These two
models explain why radio loud quasars tend to have more luminous radio
cores compared to FR\,II radio galaxies with similar extended radio
luminosities, and has prompted the use of the core dominance
parameter, $R \equiv P^\mathrm{c}_5/P^\mathrm{e}_5$ as a relative
measure of orientation. [In an attempt to have a meaningful, concise,
and consistent notation throughout this paper, we employ the
superscripts c, e, and t to refer to the core, extended, and total
radio emission (power, $P$, or flux, $S$) respectively, and subscripts
to indicate the rest-frame frequency (in GHz).]

Recently, Qin et al.\ (\cite{qin}, hereafter Q98) investigated the
core and extended radio properties of four samples of quasars from the
literature. They found that in core-dominated quasars, the extended
radio power, $P^\mathrm{e}$, was inversely correlated with $R$, while
the core radio power, $P^\mathrm{c}$, was not correlated at all.  In
lobe-dominated quasars, on the other hand, $P^\mathrm{e}$ was
uncorrelated while $P^\mathrm{c}$ was correlated with $R$. They
claimed that these correlations were in contradiction with the unified
schemes and beaming models, and proposed instead that the core
emission in core-dominated quasars is unbeamed and isotropic. In this
\emph{Letter}, we show that the results found by Q98 \emph{are} in
agreement with beaming models, given the selection criteria of the
samples they studied, and that core-dominated quasars do not comprise
a separate class of active galactic nucleus. In \S\ref{sec:beam} we
explain the reasons for the correlations observed (or unobserved) by
Q98, and in \S\ref{sec:sim} we compare the predictions of Monte Carlo
simulations of extragalactic radio sources with the data from the
samples analyzed by Q98.

\section{Beaming models}
\label{sec:beam}

In beaming models, it is assumed that all radio sources are
intrinsically steep-spectrum, and flat-spectrum sources are merely
those where the core is boosted sufficiently strongly to dominate the
total flux over the observed frequency range. The extended radio
luminosity of \emph{all} radio sources therefore follows the radio
luminosity function for steep-spectrum sources, in which the number of
sources above a given luminosity decreases sharply with luminosity.

\subsection{Core-dominated quasars}

We first consider core-dominated quasars (CDQs). Q98 believe that
$P^\mathrm{c}$ should be correlated with $R$ because those sources
with the most luminous cores should be those with the strongest
beaming. This is obviously true at some level, since both Punsly's
(1995) and Murphy, Browne \& Perley's (1993) samples, which are
selected on core strength, contain only strongly-beamed sources.
However, we show by considering the situation in more detail that the
presence or absence of correlations is controlled by selection
effects, and that this simplistic belief is too na\"{\i}ve.

Beaming models predict that the observed core dominance parameter
should depend on the jet speed, $\beta c$, the viewing angle,
$\theta$, and the transverse (unbeamed) core dominance parameter,
$R_\mathrm{T}$, as
\[
R = \frac{1}{2}R_\mathrm{T} (1 - \beta \cos \theta)^{-2} +
\frac{1}{2}R_\mathrm{T} (1 + \beta \cos \theta)^{-2}
\]
(e.g.\ Orr \& Browne \cite{orr}), where the second term can be ignored
for small values of $\theta$. Obviously, the orientation of any given
source is independent of its extended radio luminosity, and let us
assume for the moment that $\beta$ and $R_\mathrm{T}$ are also
independent of $P^\mathrm{e}$. In this case, the extended radio
luminosity function (ERLF) is separable in some form $\rho = f(P,z)
g(R)$ and sources with a given value of $R$ follow the same ERLF as
the general source population. Let us examine Punsly's (\cite{punsly})
sample, which is Q98's Sample~1. This consists of RLQs with core
powers (integrated over $10\,\mathrm{MHz} < \nu < 250\,\mathrm{GHz}$)
above $10^{39}$\,W.  Consider those sources with, say $R \approx
10^3$; they need to have $P^\mathrm{e} > 10^{36}$\,W to satisfy the
selection criterion. The shape of the ERLF is such that about
three-quarters of all such sources will lie within a factor two of
this limit, and only about 1\% will exceed it by an order of
magnitude. Most sources with $R \approx 10^3$ will therefore have
$P^\mathrm{e} \approx 10^{36}$\,W. If one now considers sources with
$R \approx 10^2$, the same arguments indicate that these sources will
have $P^\mathrm{e} \approx 10^{37}$\,W. Therefore the extended radio
power will be strongly inversely correlated with $R$, which is exactly
what is observed in Punsly's sample. Since the core luminosity is
merely $P^\mathrm{c} \equiv R P^\mathrm{e}$, the core power will be
uncorrelated with $R$. Exactly the same argument applies for Sample~2
(from Murphy et al.\ \cite{murphy}), which is a sample selected on
core flux, rather than luminosity. It is the fact that these samples
are selected on core strength that causes the observed correlations,
and not the fact that the sources are CDQs.

The maximum observable core dominance parameter is $R_\mathrm{max}
\approx \frac{1}{2} R_\mathrm{T} (1 - \beta)^{-2}$. For typical
values of $\beta = 0.99$ ($\Gamma = 7$, e.g.\ Guerra \& Daly
\cite{guerra}) and $\log R_\mathrm{T} = -2.5$ (at 5\,GHz, Simpson
\cite{simpson1}; see also Morganti et al.\ \cite{morganti}), this is
about 20 using the defintion and canonical radio spectra of Punsly
(\cite{punsly}). To exceed this value by nearly two orders of
magnitude (as some of Punsly's sources do), a source must either have
an unusually fast jet ($\beta = 0.999$, $\Gamma = 22$) or a much
larger than average $R_\mathrm{T}$. These parameters will therefore be
correlated with the observed core dominance, and so could an
additional correlation with extended radio power nullify the above
argument?

Suppose that $\beta$ is correlated with the extended radio luminosity.
This is not unreasonable since the most powerful radio sources might
also produce the fastest jets. If so, sources with large $R$ are more
likely to be drawn from the high-luminosity end of the ERLF. On the
other hand, $R_\mathrm{T}$ might be anticorrelated with extended radio
luminosity, since the lobe luminosity of a radio source decreases
throughout its lifetime despite a constant jet power (e.g.\ Baldwin
\cite{baldwin}). So sources with large $R$ might preferentially be
older sources with lower extended radio luminosities. The fact that
the anticorrelation between $\log P^\mathrm{e}$ and $\log R$ appears
not to alter its slope at large $R$ indicates that either both these
effects are weak, or that they have approximately the same strength
and cancel each other out.

\subsection{Lobe-dominated quasars}

We now turn to lobe-dominated quasars (LDQs). Hooimeyer et al.'s
(\cite{hooimeyer}) sample of quasars (Q98's Sample~3) is selected on a
minimum angular size, thereby requiring measurable emission from the
radio lobes and excluding the most strongly-boosted, pole-on sources,
but also has a bright core radio flux limit ($S^\mathrm{c}_5 >
0.1$\,Jy), which means that it will include many CDQs. In addition,
the sample is drawn from the Hewitt \& Burbidge (\cite{hewitt})
optical quasar catalogue and is therefore subject to uncertain
selection effects.  The fact that in reality it is a mix of CDQs and
LDQs can serve to explain why it displays both correlations, but at a
much weaker level than in samples of CDQs or LDQs alone. We do not
consider it further.

Instead, we turn to the well-defined Sample~4, from Hough \& Readhead
(\cite{hough}) which contains all double-lobed quasars from 3CRR
(Laing, Riley \& Longair \cite{laing}) whose cores were bright enough
to be mapped with the Mark~III VLBI system. All sources therefore have
$S_{0.178}^\mathrm{t} > 10.9$\,Jy and $S_5^\mathrm{c} > 5$\,mJy.
Unlike the core-selected samples, the total flux at the selection
frequency (178\,MHz) has a negligible contribution from the
flat-spectrum core, and therefore $S_{0.178}^\mathrm{t}$ has no
dependence on $R$. Neither does $P_{0.178}^\mathrm{t}$, which can be
assumed to be the extended luminosity alone. The spectral index of the
lobes is not expected to be physically related to core dominance, and
we confirm that this is so in the sample of radio galaxies from
Hardcastle et al.\ (\cite{hardcastle}) by using the generalized
Spearman's rank order correlation coefficient (Isobe, Feigelson \&
Nelson \cite{isobe}), where we find that the two quantities are
uncorrelated at greater than 90\% confidence. Therefore,
$P_5^\mathrm{e}$ will be uncorrelated with $R$. On the other hand,
$P_5^\mathrm{c} = R \, P_5^\mathrm{e}$, and should therefore be
clearly correlated with $R$, with the linear regression line having a
slope of unity, in line with observations. Again, it is not the nature
of the sources (LDQs) which produces the result, but the selection
criteria.

\section{Monte Carlo simulations}
\label{sec:sim}

In order to test whether the above arguments can explain the
observations in a quantitative manner, we perform simulations of radio
sources. We start with the steep spectrum radio luminosity function
(RLF1) of Dunlop \& Peacock (\cite{dunlop}), which we assume to be
equivalent to the extended luminosity of all radio sources, and
consider only those sources with luminosities above the Fanaroff-Riley
break (i.e.\ FR\,II sources; Fanaroff \& Riley \cite{fanaroff}), since
it is only these which constitute the parent population of quasars
(e.g.\ Urry \& Padovani \cite{urry}). We assume that the sources are
randomly oriented with respect to the observer, and that their $\log
R_\mathrm{T}$ values are drawn from a Normal distribution with mean
$-2.43$ and standard deviation 0.50. This distribution provides a good
fit to the observed values of $R$ for FR\,II radio galaxies and
quasars from Laing et al.\ (\cite{laing}) with $z < 0.43$, after
applying the Orr \& Browne (\cite{orr}) beaming model (Simpson et al.\
\cite{simpson2}; see also Simpson \cite{simpson1}). It is assumed that
the radio cores have $\alpha = 0$ and the extended emission has
$\alpha = 0.7$ ($S_\nu \propto \nu^{-\alpha}$). A jet speed $\beta =
0.99$ is assumed for the purposes of simulating Doppler boosting. We
then subject the ensemble of sources to the selection criteria of the
samples studied by Q98, and compare our regression and correlation
analysis with theirs. As the samples of Q98 invariably have a number
of sources with incomplete information that prevent them being
included in Q98's analysis, we also randomly exclude sources from our
complete samples so as to have the same sample size. This ensures that
the same level of randomness should exist in our simulated samples as
in the real data. All our simulations assume $H_0 =
50$\,km\,s$^{-1}$\,Mpc$^{-1}$, $q_0 = 0.5$ and $\Lambda = 0$, although
the significance of the correlations we find do not depend on our
choice of cosmology.

\subsection{Sample 1}

\begin{figure*}
\resizebox{\hsize}{!}{\includegraphics{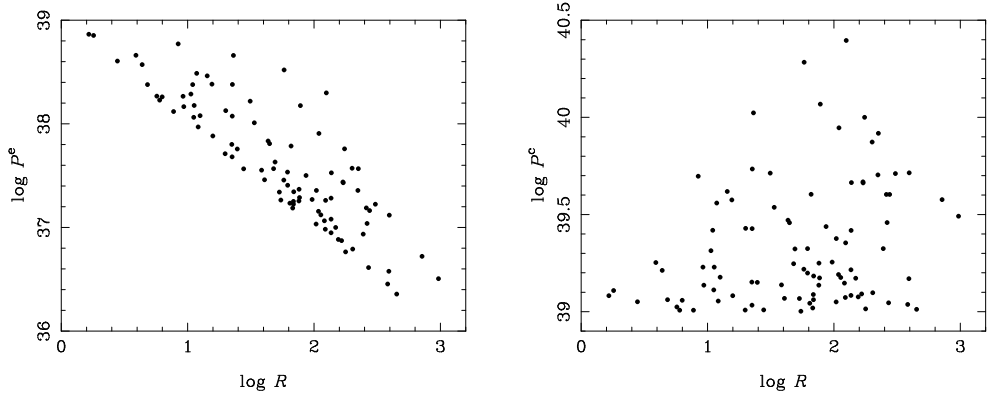}}
\caption[]{Plots of extended and core radio luminosity (integrated
from 10\,MHz--250\,GHz) \emph{vs} the ratio of these quantities, for
a simulated dataset with the same selection criteria as that of Punsly
(\cite{punsly}).}
\label{fig:sample1}
\end{figure*}

Sample~1 (Punsly \cite{punsly}) consists of quasars with core
luminosities, integrated over the range $10\,\mathrm{MHz} < \nu <
250\,\mathrm{GHz}$, greater than $10^{39}$\,W. Adopting the typical core
spectrum assumed by Punsly, this corresponds to $P_5^\mathrm{c} > 8
\times 10^{27}$\,W\,Hz$^{-1}$, which we adopt as our selection
criterion.  In Figure~\ref{fig:sample1} we plot the extended and core
luminosities against the core dominance parameter.

\subsection{Sample 2}

\begin{figure*}
\resizebox{\hsize}{!}{\includegraphics{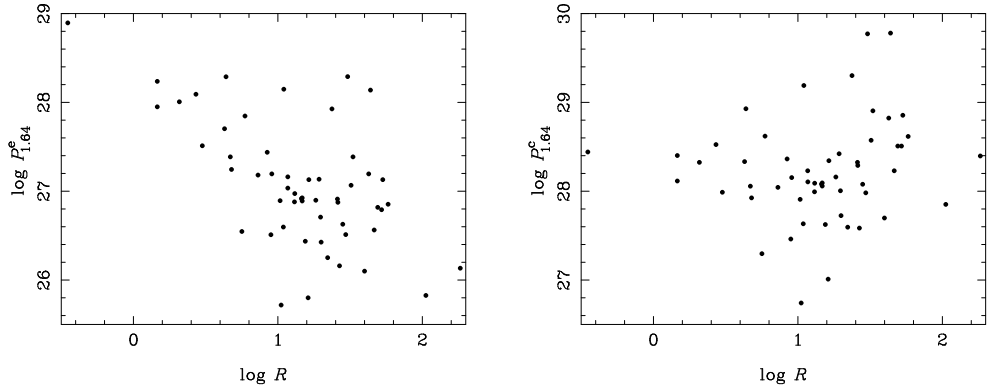}}
\caption[]{Plots of extended and core radio luminosity (at 1.64\,GHz
in the rest frame of the source), \emph{vs} the ratio of these
quantities for a simulated dataset with the same selection criteria as
that of Murphy et al.\ (\cite{murphy}).}
\label{fig:sample2}
\end{figure*}

Sample~2 (Murphy et al.\ \cite{murphy}) contains sources with core
fluxes $S_5^\mathrm{c} \geq 1$\,Jy and spectral indices (between 1.4
and 5\,GHz) $\alpha \leq 0.5$ which are identified with quasars. This
last criterion will exclude nearby sources since they would be
classified as (broad line) radio galaxies. Examining Murphy et al.'s
sample, we elect to discard all sources from our dataset with $z <
0.25$ to simulate this. There are in fact only about three sources at
such low redshifts in a typical dataset, so the results are not
strongly affected. The simulated datasets are shown in
Figure~\ref{fig:sample2}.

\subsection{Sample 3}

As discussed earlier, Sample~3 (Hooimeyer et al.\ \cite{hooimeyer})
has uncertain selection criteria, and so we make no attempt to
simulate it.

\subsection{Sample 4}

\begin{figure*}
\resizebox{\hsize}{!}{\includegraphics{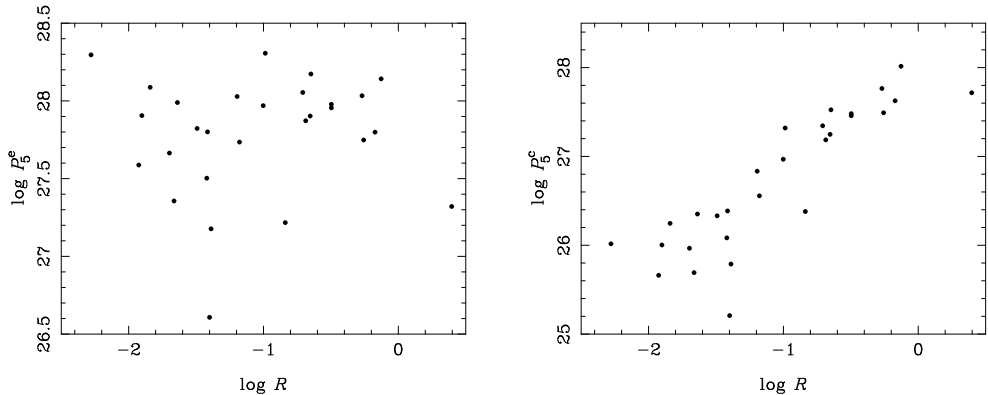}}
\caption[]{Plots of extended and core radio luminosity (at rest-frame
5\,GHz) \emph{vs} the ratio of these quantities for a simulated
dataset with the same selection criteria as that of Hough \& Readhead
(\cite{hough}).}
\label{fig:sample4}
\end{figure*}

Sample~4 (Hough \& Readhead \cite{hough}) consists of sources with
$P_{0.178}^\mathrm{t} > 10.9$\,Jy and $S_5^\mathrm{c} > 5$\,mJy that
are classified as quasars. Since this is a much fainter core flux
limit than Sample~2, we cannot just apply a redshift cutoff, since
many distant narrow line radio galaxies will have sufficiently bright
cores. Instead we require (in addition to $z > 0.25$), that $\theta <
45\degr$ (see Barthel \cite{barthel}), which produces the sample
displayed in Figure~\ref{fig:sample4}.

\begin{table}
\caption[]{Correlation coefficients, $r$, and slopes of the best-fit
regression lines, $a$, for the three samples modelled. For each
sample, the top line contains the results from the real data
determined by Q98, and the bottom line contains the results from the
simulated data.}
\label{tab:results}
\begin{center}
\begin{tabular}{ccrrrr}
\hline
Sample & $N$ & \multicolumn{1}{c}{$r_{P^\mathrm{e},R}$} &
\multicolumn{1}{c}{$r_{P^\mathrm{c},R}$} &
\multicolumn{1}{c}{$a_{P^\mathrm{e},R}$} &
\multicolumn{1}{c}{$a_{P^\mathrm{c},R}$} \\
\hline
1 & 94 & $-0.899$ & 0.196 & $-0.91 \pm 0.05$ & $0.09 \pm 0.05$ \\
  &    & $-0.864$ & 0.257 & $-0.87 \pm 0.10$ & $0.13 \pm 0.10$ \\
2 & 54 & $-0.739$ & 0.150 & $-0.88 \pm 0.11$ & $0.12 \pm 0.11$ \\
  &    & $-0.591$ & 0.136 & $-0.84 \pm 0.10$ & $0.16 \pm 0.10$ \\
4 & 28 & $-0.286$ & 0.848 & $-0.16 \pm 0.10$ & $0.84 \pm 0.10$ \\
  &    &   0.056  & 0.873 &  $0.03 \pm 0.10$ & $1.03 \pm 0.10$ \\
\hline
\end{tabular}
\end{center}
\end{table}

\subsection{Discussion}

Table~\ref{tab:results} presents the results of correlation and
regression analysis on the simulated datasets, and clearly shows that
the simulated datasets display exactly the same correlations as the
real data. Whether it is core or extended radio luminosity that is
correlated with core dominance is therefore not due to some physical
difference between CDQs and LDQs, but merely an effect caused by the
sample selection criteria, as discussed in \S\ref{sec:beam}. We note
that our simulated data underproduces the number of highly
core-dominated ($R
\mbox{\raisebox{-.5ex}{$\stackrel{\textstyle>}{\sim}$}} 10^2$)
quasars, compared to observations. This is presumably a result of our
not having incorporated a range of jet speeds into our model, since
with a single speed ($\beta = 0.99$) such sources require a viewing
angle close to the line of sight \emph{and} a large value of
$R_\mathrm{T}$. If the jet speed is a third random variable then such
sources can be created if only two out of the three variables have
extreme values, and this would therefore increase their frequency in
our simulated Universe.

\section{Summary}

We have investigated the claim by Qin et al.\ (\cite{qin}) that the
observed correlations between core dominance and core or extended
radio luminosity in samples of core- and lobe-dominated quasars are in
contradiction with the beaming models which form a part of the unified
scheme for extragalactic radio sources. We have argued that these
correlations are actually \emph{expected} to arise from the beaming
models as a natural consequence of the steepness of the (extended)
radio luminosity function and the selection criteria of the samples.
We have supported our claim by simulating quasar samples using the
observed radio luminosity function and the statistics of beaming, and
find that our simulated data produce results which are
indistinguishable from those produced by the real data on which they
are based.

\begin{acknowledgements}

It is a pleasure to thank Katherine Blundell for reading the
manuscript prior to submission and providing insightful comments.

\end{acknowledgements}


\begin{thebibliography}{}
\bibitem[1982]{baldwin}
Baldwin, J.E. 1982, in Extragalactic Radio Sources, ed.\ D.S. Heeshen
\& C.M. Wade (Dordecht: Reidel), 21
\bibitem[1989]{barthel}
Barthel, P.D. 1989, ApJ, 336, 606
\bibitem[1990]{dunlop}
Dunlop, J.S., Peacock, J.A. 1990, MNRAS, 247, 19
\bibitem[1974]{fanaroff}
Fanaroff, B.L., Riley, J.M. 1974, MNRAS, 167, 31{\sc p}
\bibitem[1997]{guerra}
Guerra, E.J., Daly, R.A. 1997, ApJ, 491, 483
\bibitem[1998]{hardcastle}
Hardcastle, M.J., Alexander, P., Pooley, G.G., Riley, J.M., 1998,
MNRAS, 296, 445
\bibitem[1987]{hewitt}
Hewitt, A., Burbidge, G. 1987, ApJS, 63, 1
\bibitem[1992]{hooimeyer}
Hooimeyer, J.R.A., Barthel, P.D., Schilizzi, R.T., Miley, G.K.
1992, A\&A, 261, 18
\bibitem[1989]{hough}
Hough, D.H., Readhead, A.C.S. 1989, AJ, 98, 1208
\bibitem[1986]{isobe}
Isobe, T., Feigelson, E.D., Nelson, P.I., 1986, ApJ, 306, 490
\bibitem[1983]{laing}
Laing, R.A., Riley, J.M., Longair, M.S. 1983, MNRAS, 204, 151
\bibitem[1997]{morganti}
Morganti, R., Oosterloo, T.A., Reynolds, J.E., Tadhunter, C.N.,
Migenes, V., 1997, MNRAS, 284, 541
\bibitem[1992]{murphy}
Murphy, D.W., Browne, I.W.A., Perley, R.A. 1992, MNRAS, 264, 298
\bibitem[1982]{orr}
Orr, M.J.L., Browne, I.W.A. 1982, MNRAS, 200, 1067
\bibitem[1995]{punsly}
Punsly, B. 1995, AJ, 109, 1555
\bibitem[1998]{qin}
Qin, Y.P., Xie, G.Z., Bai, J.M., Fan, J.H., Zheng, X.T., Wu, S.M.
1998 , A\&A, 333, 790 (Q98)
\bibitem[1996]{simpson1}
Simpson, C., 1996, VA, 40, 57
\bibitem[1998]{simpson2}
Simpson, C., et al., 1998, in preparation
\bibitem[1995]{urry}
Urry, C.M., Padovani, P., 1995, PASP, 107, 803
\end{thebibliography}
\end{document}